# APPLICATION OF GAMMACHIRP AUDITORY FILTER AS A CONTINUOUS WAVELET ANALYSIS

Lotfi Salhi[1] and Kais Ouni[2]

[1]Department of Physics, Sciences Faculty of Tunis (FST), University of Tunis ELManar, Tunisia
lotfi.salhi@laposte.com

[2]Systems and Signal Processing Laboratory, National School of Engineers of Tunis (ENIT), University of Tunis ELManar, Tunisia
kais.ouni@enit.rnu.tn

## ABSTRACT

*This paper presents a new method on the use of the gammachirp auditory filter based on a continuous wavelet analysis. The gammachirp auditory filter is designed to provide a spectrum reflecting the spectral properties of the cochlea, which is responsible for frequency analysis in the human auditory system.*
*The impulse response of the theoretical gammachirp auditory filter that has been developed by Irino and Patterson can be used as the kernel for wavelet transform which approximates the frequency response of the cochlea. This study implements the gammachirp auditory filter described by Irino as an analytical wavelet and examines its application to a different speech signals.*
*The obtained results will be compared with those obtained by two other predefined wavelet families that are Morlet and Mexican Hat. The results show that the gammachirp wavelet family gives results that are comparable to ones obtained by Morlet and Mexican Hat wavelet family.*

## KEYWORDS

*Gammachirp, Wavelet Transform, Cochlear Filter, speech Processing*

## 1. INTRODUCTION

In order to understand the auditory human system, it is necessary to approach some theoretical notions of our auditory organ, in particular the behavior of the internal ear according to the frequency and according to the resonant level.

The sounds arrive to the pavilion of the ear, where they are directed towards drives in its auditory external. To the extremity of this channel, they exercise a pressure on the membrane of the eardrum, which starts vibrating to the same frequency those them. The ossicles of the middle ear, interdependent of the eardrum by the hammer, also enter in vibration, assuring the transmission of the soundwave thus until the cochlea. The resonant vibration arrives to the cochlea by the oval window, separation membrane between the stirrup, last ossicle of the middle ear, and the perilymphe of the vestibular rail. The endolymphe of the cochlear channel vibrates then on its turn and drag the basilar membrane. The stenocils, agitated by the liquidizes movements, transforms the acoustic vibration in potential of action (nervous messages); these last are transmitted to the brain through the intermediary of the cochlear nerve [5][6][7].

These mechanisms of displacement on any point of the basilar membrane, can begins viewing like a signal of exit of a pass - strip filter whose frequency answer has its pick of resonance to a frequency that is characteristic of its position on the basilar membrane. The cochlea can be seen like a bench of pass - strip filters [9][10][11].

To simulate the behavior of these filters, several models have been proposed. Thus, one tries to succeed to an analysis of the speech signals more faithful to the natural process in the progress of a signal since its source until the sound arrived to the brain. By put these models, one mentions the model gammachirp that has been proposed by Irino & Patterson [1][2][3].

While being based on the impulsional answer of this filter type, it comes the idea to implement as family of wavelet of which the function of the wavelet mother is the one of this one.

## 2. GAMMACHIRP AUDITORY FILTER

Within the cochlea, sound waves travel through a fluid and excite small hair cells along the basilar membrane. High frequency tones excite hair cells near the oval window whereas low frequency tones affect hair cells near the end. Several models have been proposed to simulate the working of the cochlear filter. Seen as its temporal-specter properties, the gammachirp filter underwent a good success in the psychoacoustic research. Indeed, globally it answers the requirements and the complexities of the cochlear filter. In addition to its good approximation in the psycho acoustical model, it possesses a temporal-specter optimization of the human auditory filter [4].

The notion of the wavelet transform possesses a big importance in the signal treatment domain and in the speech analysis. it comes the idea to exploit the theory of the wavelet transform in the implementation of an auditory model of the cochlear filter proposed by Irino and Patterson [2][3] that is the Gammachirp filter.

One is going to be interested to the survey of this filter type and to its implementation as a wavelet under the Matlab flat forms. The impulsional answer of the gammachirp filter is given by the following function [1][2][3]:

$$g(t) = \lambda_n t^{n-1} e^{-2\pi b ERB(f_0)t} e^{j(2\pi f_0 t + c \ln(t) + \phi)} \quad (1)$$

With: $t > 0$
  $n$ : a whole positive defining the order of the corresponding filter.
  $f_0$ : the modulation frequency of the gamma function.
  $\varphi$ : the original phase,
  $\lambda_n$ an amplitude normalization parameter.
  $ERB(f_0)$: Equivalent Rectangulaire Bandwith

The gamma envelope and the frequency glide were originally proposed to characterize the revcor-data of basilar membrane motion (BMM). The gammachirp is consistent with basic physiological data. This model provides excellent fit to human masking data. The most advantage is that the filter can handle both of physiological and psychoacoustical data within a unified framework.

When $c=0$, the chirp term, $c \ln(t)$, vanishes and this equation represents the complex impulse response of the gammatone that has the envelope of a gamma distribution function and its carrier is a sinusoid at frequency $f_r$. Accordingly, the gammachirp is an extension of the gammatone with a frequency modulation term.

The frequency responses of the gammachirp filters, as seen in Figure 1, are asymmetric and exhibit a sharp drop off on the high frequency side of the center frequency. This corresponds well to auditory filter shapes derived from masking data.

The amplitude spectrum of the gammachirp can be written in terms of the gammatone as:

$$G(f) = \int_0^{+\infty} g(t)e^{-j2\pi ft}dt = \int_0^{+\infty} \lambda_n t^{n-1+jc}e^{-2\pi\beta t}e^{j2\pi f_0 t+j\phi}e^{-j2\pi ft}dt \quad (2)$$

Where $\beta = b.ERB(f_0)$ so

$$G(f) = \frac{\alpha}{(2\pi\beta)^{n+jc}\left[1+j\frac{(f-f_0)}{\alpha}\right]^{n+jc}} \quad (3)$$

Where $\alpha = \lambda_n \Gamma(n+jc)e^{j\phi}$ and $\Gamma(n+jc)e^{j\phi}$ is the complex distribution of gamma.

$$G_C(f) = \left[\alpha . \frac{1}{\left\{2\pi\sqrt{\beta^2+(f-f_0)^2}\right\}^n} . e^{-jn\theta}\right] . \left[e^{c\theta} . e^{-jc\ln\left\{2\pi\sqrt{\beta^2+(f-f_0)^2}\right\}}\right] \quad (4)$$

where $\theta = arctg(\frac{f-f_0}{\beta})$

The amplitude spectrum of the gammachirp can be written in terms of the gammatone as:

$$|G_C(f)| = \frac{1}{\left\{2\pi\sqrt{\beta^2+(f-f_0)^2}\right\}^n} . e^{c\theta} = a_\Gamma(c)|G_T(f)|.e^{c\theta} \quad (5)$$

Where $GC(f)$ is the Fourier transform of the gammachirp function, $GT(f)$ is the Fourier transform of the corresponding gammatone function, $c$ is the chirp parameter, $a_¡(c)$ is a gain factor which depends on $c$.

An example of the impulse response of the Gammachirp (GC) filter and its spectrum is given by the following figure. The gammatone is indicated by (GT) and the asymmetric function by (AF).

This decomposition, which was shown by Irino in [2], is beneficial because it allows the gammachirp to be expressed as the cascade of a gammatone filter, $GT(f)$, with an asymmetric compensation filter, $e^{c\theta}$. Figure 1 shows the frame-work for this cascade approach. The spectrum of the overall filter can then be made level-dependent by making the parameters of the asymmetric component depend on the input stimulus level.

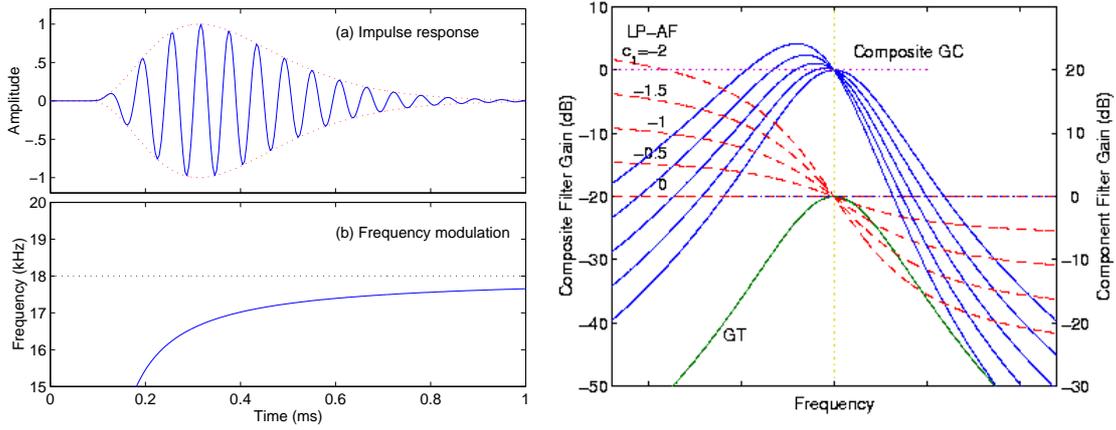

Figure 1.  Impulse response of the Gammachirp auditory filter and its spectrum

The figure 2 shows a filterbank structure for the physiological gammachirp. It is a cascade of three filterbanks: a gammatone filterbank, a lowpass-AC filterbank, and a highpass-AC filterbank.

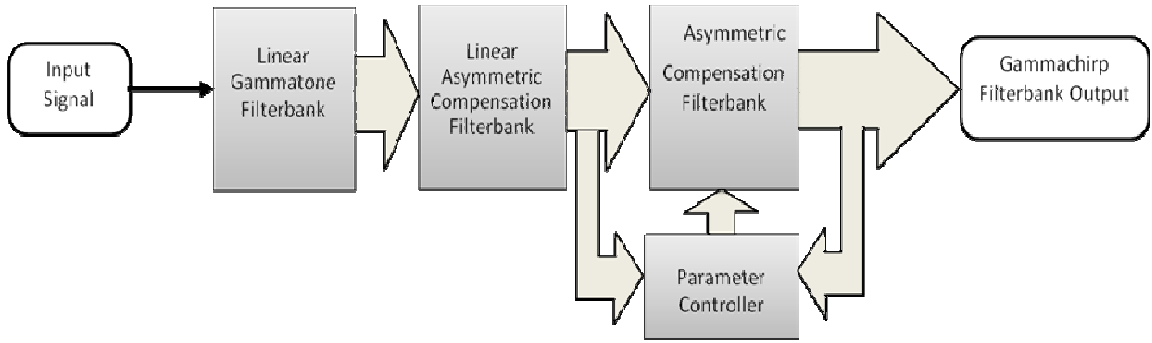

Figure 2.  Structure of the Gammachirp filterbank

## 3. APPLICATION OF THE GAMMACHIRP AS ANALYTICAL WAVELET

### 3.1. Gammachirp wavelet

The gammachirp function can be considered like wavelet function and constitute a basis of wavelets thus on the what be project all input signal, it is necessary that it verifies some conditions that are necessary to achieve this transformation. [12][13][14]

Indeed it must verify these two conditions:

1- The wavelet function must be a finished energy:

$$\|g\|^2 = \langle g, g^* \rangle = \int_{-\infty}^{+\infty} |g(t)|^2 dt = \frac{\lambda_n^2 \, \Gamma(2n-1)}{(4\pi\beta)^{2n-1}} = \frac{\lambda_n^2 \, (2n-1)!}{(4\pi\beta)^{2n-1}} \tag{6}$$

$\|g\|^2 = 1$ if $\lambda_n = \sqrt{\dfrac{(4\pi\beta)^{2n-1}}{(2n-1)!}}$ which define the filter of normalized energy.

2- The wavelet function must verify the admissibility condition:

$$C_g = \int_0^{+\infty} \frac{|G(f)|^2}{f} df < +\infty \tag{7}$$

If the condition (7) is satisfied by the function G, then it must satisfy two other conditions:

- The mean function g is zero: $G(0) = \int_{-\infty}^{+\infty} g(t)dt = 0$

- The function G (f) is continuously differentiable

To implement the Gammachirp function g as wavelet mother, one constructs a basis of wavelets then girls $g_{a,b}$ and this as dilating g by factor 'a' and while relocating it of a parameter 'b'.

$$g_{a,b}(t) = \frac{1}{\sqrt{a}} g\left(\frac{t-b}{a}\right) \tag{8}$$

Studies have been achieved on the gammachirp function [4], show that the Gammachirp function that is an amplitude-modulated window by the frequency $f_0$ and modulated in phase by the c parameter, can be considered like roughly analytic wavelet. it is of finished energy and it verifies the condition of admissibility.

This wavelet possesses the following properties: [4] it is not symmetrical, it is non orthographic and it doesn't present scale function. The implementation convenient of the wavelet requires a discretization of the dilation parameters and the one of transfer.

One takes $a = a_0^m$ et $b = kb_0 a_0^m$ ($k$ et $m \in \mathbb{Z}$).

She gotten wavelets girls have for expression:

$$g_{m,k}(t) = g_{(a_0^m, kb_0 a_0^m)}(t) = a_0^{-m/2} g(a_0^{-m} t - kb_0) \tag{9}$$

The results gotten based on the previous works [4] shows that the value 1000 Hz are the one most compatible as central frequency of the Gammachirp function. Otherwise our work will be based on the choice of a Gammachirp wavelet centered at the frequency 1000 Hz. For this frequency range, the gammachirp filter can be considered as an approximately analytical wavelet.

### 3.1. Other predefined wavelets

There are many other wavelets that have the same properties like gammachirp wavelet, such as derivatives of Gaussian functions, Morlet wavelet, Mexican hat wavelet, etc. The selection of a particular function is based upon criteria set by the use of the function. Each individual function is known as a mother wavelet and compressed and dilated versions of this mother are used throughout the wavelet analysis process. For example:

- ***Derivative of Gaussian wavelet*** :  $gaus(x,n) = Cn * diff(e^{-x^2}, n)$

    Where *Cn* is a normalized factor and *diff* is a differential at the order *n*

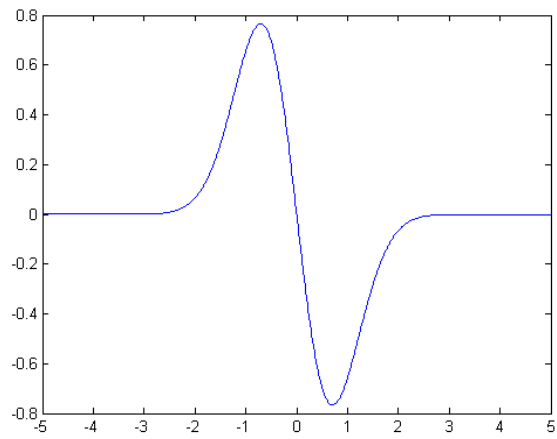

Figure 3. Gaussian wavelet at the first order

- ***Morlet wavelet***: $morl(x) = \dfrac{1}{\sqrt{2\pi}} \cos(w_0 x) e^{-\frac{x^2}{2}}$

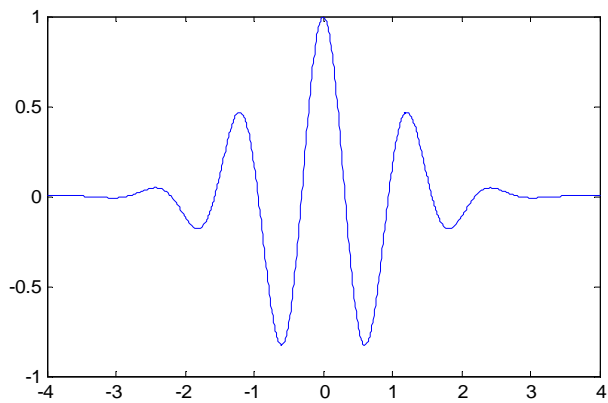

Figure 4. Morlet wavelet

- ***Mexican hat***: $mexh(x) = c(1-x^2) e^{-\frac{x^2}{2}}$   where *c* is a normalized factor.

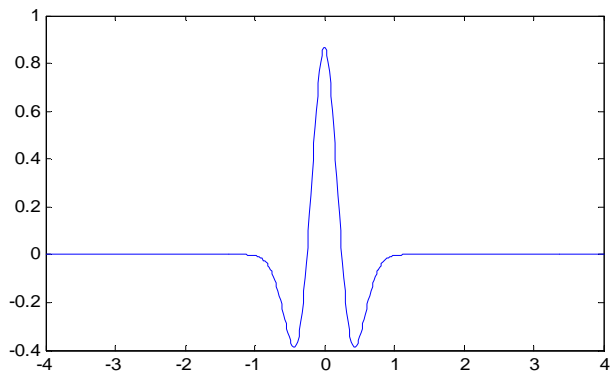

Figure 5. Mexican hat wavelet

## 4. IMPLEMENTATION AND RESULTS

To validate this implementation us applied this wavelet on three synthesized vowels like /a/, /u / and /i / of which the values of the three forming firsts is regrouped in this table:

Table 1. Pitch and formants values for each vowel

| Vowel | Pitch ( Hz) | F1 (Hz) | F2 ( Hz) | F3 (Hz) |
|---|---|---|---|---|
| /a/ | 100 | 730 | 1090 | 2440 |
| /i/ | 100 | 270 | 2290 | 3010 |
| /u/ | 100 | 300 | 870 | 2240 |

One applies the Gammachirp wavelt implemented to an input signal that is one of the utilized vowels and that are /a /, /u / and /i /. One gets the result that represents the first five levels of analysis of the signal by this type of wavelet, as well as the corresponding specters.

This analysis will be followed by a comparison with other types of wavelets as the Morlet wavelet and Mexican Hat wavelet. The gotten results are summarized in the following sections:

### 4.1. Analysis of the vowel /a /

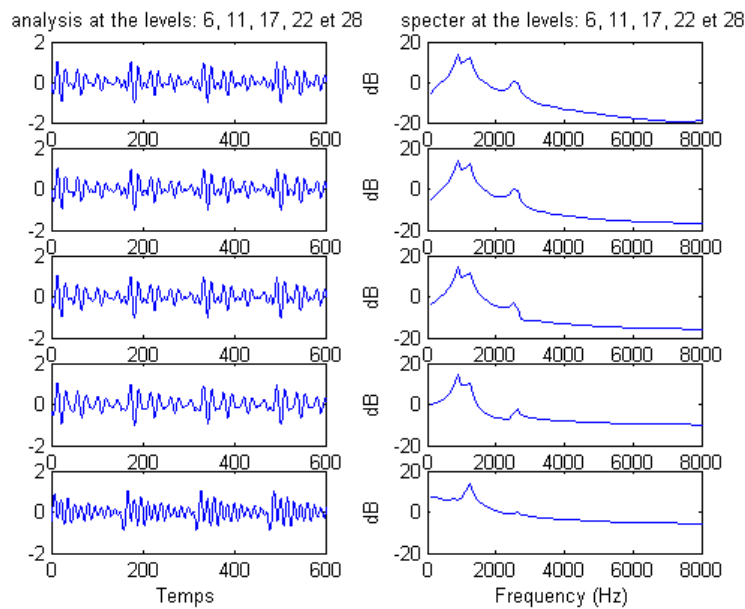

Figure 6. Analysis by the Gammachirp wavelet and the logarithmic specters correspond for the levels 6, 11, 17, 22 and 28 of the vowel /a /

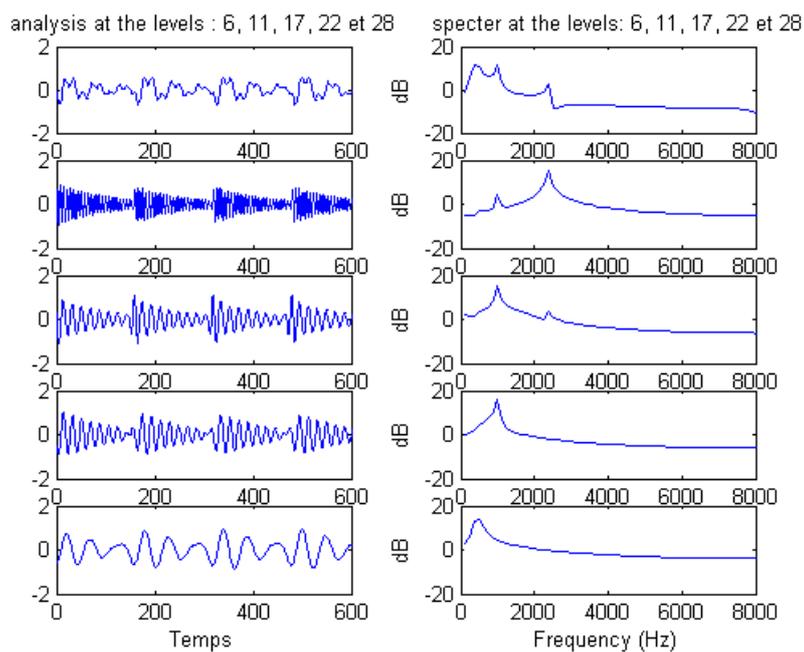

Figure 7. Application of the Morlet wavelet on the vowel /a /

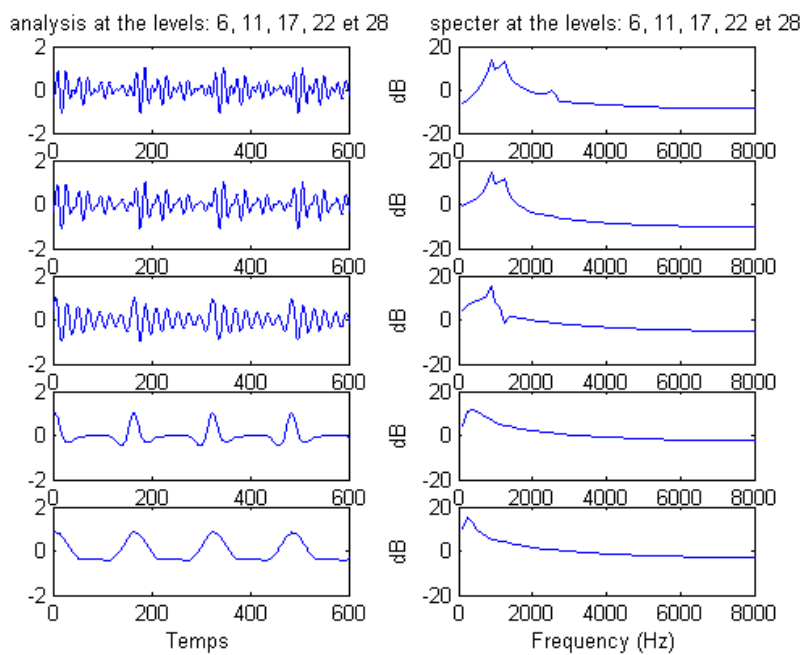

Figure 8. Application of the Mexicain Hat wavelet on the vowel /a/

Figure 6 gives a summary of the results gotten by the different wavelets:

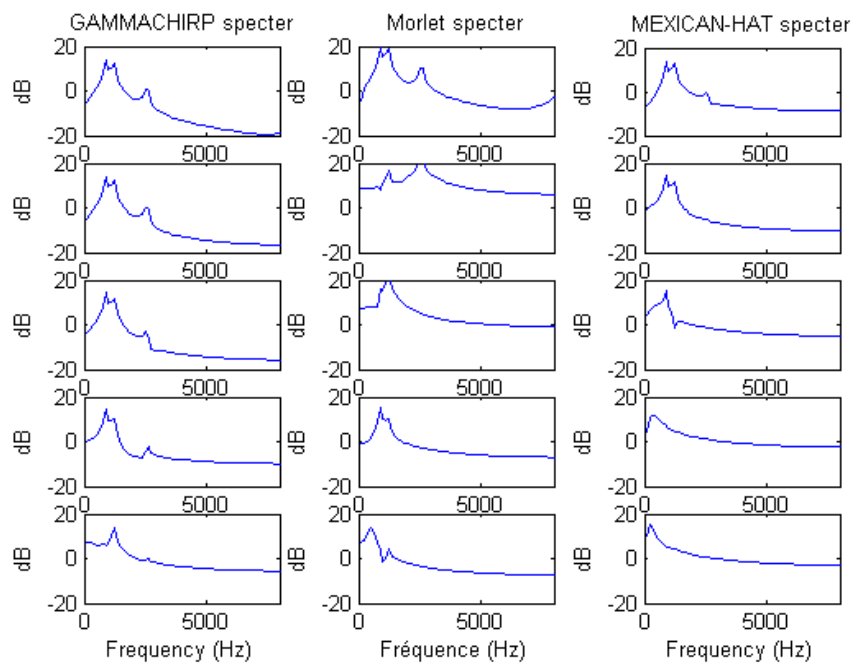

Figure 9. Comparison of the three results gotten for the vowel /a /

The gammachirp filter detects à practically the first level of analysis the three forming of the signal corresponding to the vowel /a / and it eliminates at every upper level the high frequencies and this as subdividing at every time centers it frequency by $s_0^m$ (with s0 = 1,13 and m represents the level of analysis).

Indeed, the first forming the vowel /a / corresponds to the frequency 730 Hz. It is detected in the first five levels of analysis. It is waited, since the fifth level is limited by the frequency 522Hz.

The second forming corresponds to the frequency 1090 Hz. One notices that it is only detected in the first four levels of analysis. The fourth level corresponds to the limit frequency 1087 Hz, what explains the presence of the second forming.

The third forming corresponds to the frequency 2440 Hz. One notices the attenuation of that forming begins dice the third level whose limit frequency is of 2003 Hz.

The figures (Figure 7 and Figure 8) show the results given by application of the Morlet and Mexican hat wavelets. One notices that the results found by application of the gammachirp wavelet (Figure 9) are comparable at those given by the Morlet and Mexican hat wavelets.

### 4.2. Analysis of the vowel /u /

The application of the wavelet transform as using the Gammachirp wavelet, Morlet and Mexican Hat wavelets on this vowel gives the following results:

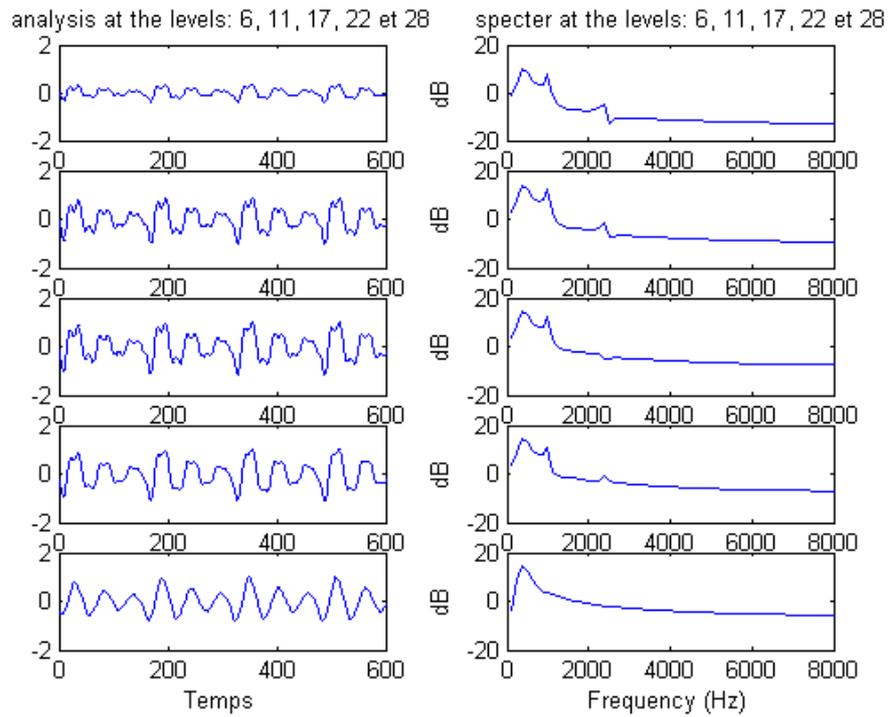

Figure 10. Application of the Gammachirp wavelet on the vowel /u/ and the logarithmic specters correspond for the levels 6, 11, 17, 22 and 28 of the vowel /a /

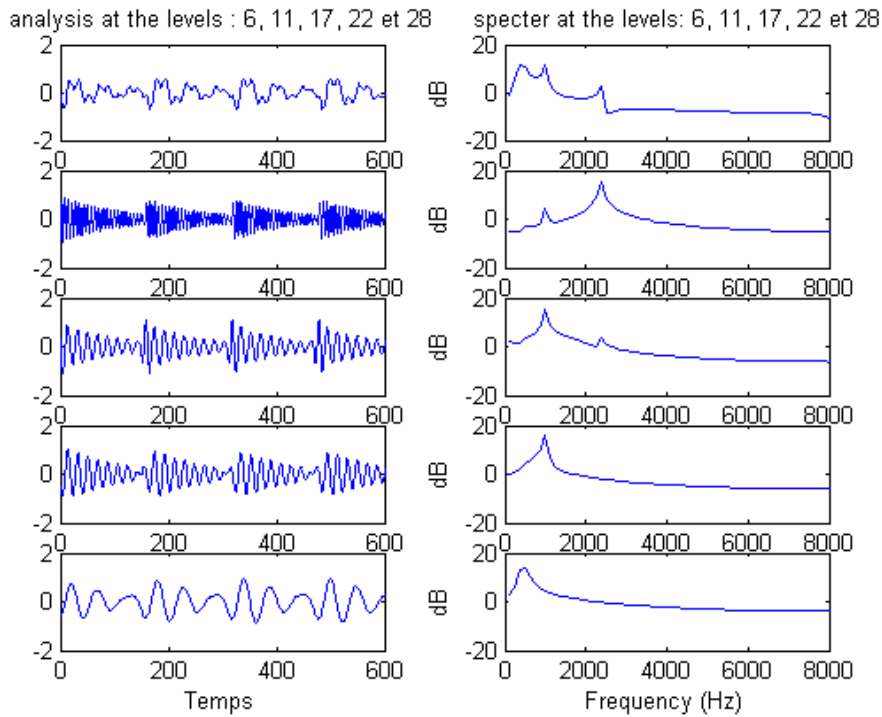

Figure 11. Application of the Morlet wavelet on the vowel /u/

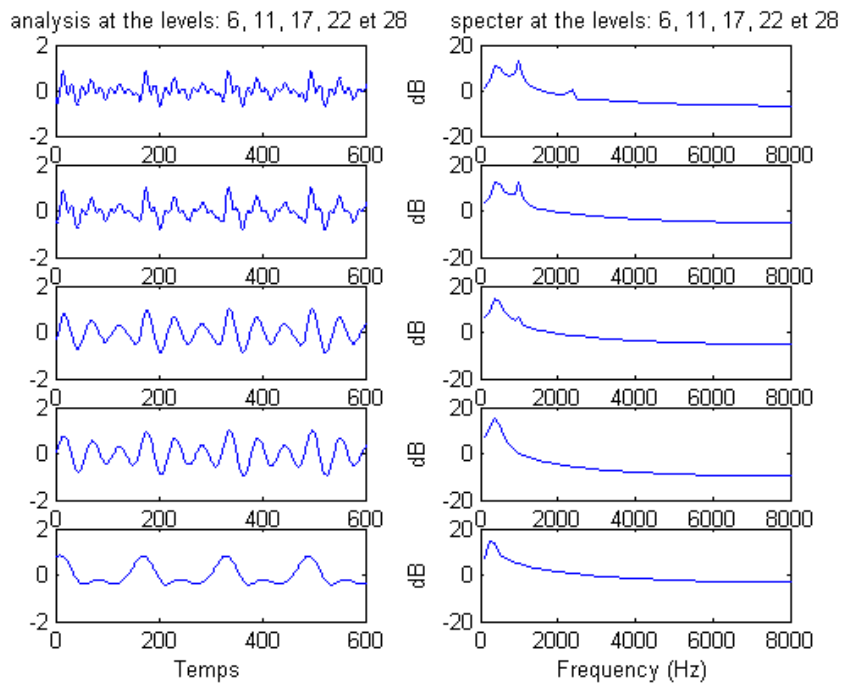

Figure 12. Application of the Mexicain Hat wavelet on the vowel /u/

Figure 10 gives a summary of the results gotten by the different wavelets:

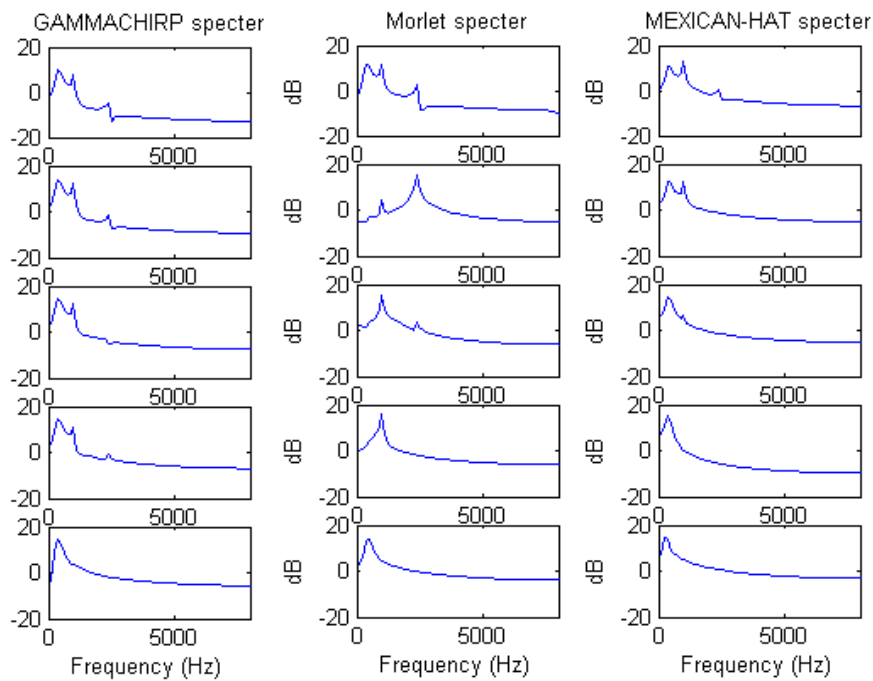

Figure 13. Comparison of the three results gotten for the vowel /u/

The first forming of the vowel /u/ corresponds to the frequency 300 Hz. He/it is detected in the five levels of analysis. It is waited, since the last level is limited by the frequency 522 Hz.
The second forming corresponds to the frequency 870 Hz. One notices that it is only detected in the first four levels of analysis (Figure 10) and this while applying the Gammachirp wavelet.

The last level corresponds to the limit frequency 522 Hz. Whereas the frequency of the second forming (870 Hz) passes this limit, what explains the presence of a weak fluctuation at the level of its specter, slightly.

The third forming corresponds to the frequency 2240 Hz. One notices the disappearance of that forming dice the third level and this practically for the three types of wavelet (Figure 11 and Figure 12). One notices by comparison with the Morlet wavelet that the Gammachirp wavelet as the Mexican hat wavelet detects better the first forming of the vowel /u /.

Thus, with regard to the detection of the three forming firsts of the vowel /u /, one notices that the results found by application of the gammachirp wavelet are comparable at those given by the Morlet and Mexican hat wavelets (Figure 13). The Gammachirp wavelet present sometimes a light improvement opposite the two firsts in the detection of the first and the third forming of the vowel /u /.

### 4.3. Analysis of the vowel /i /

The application of the wavelet transform as using the Gammachirp wavelet, Morlet and Mexican Hat wavelets on this vowel gives the following results:

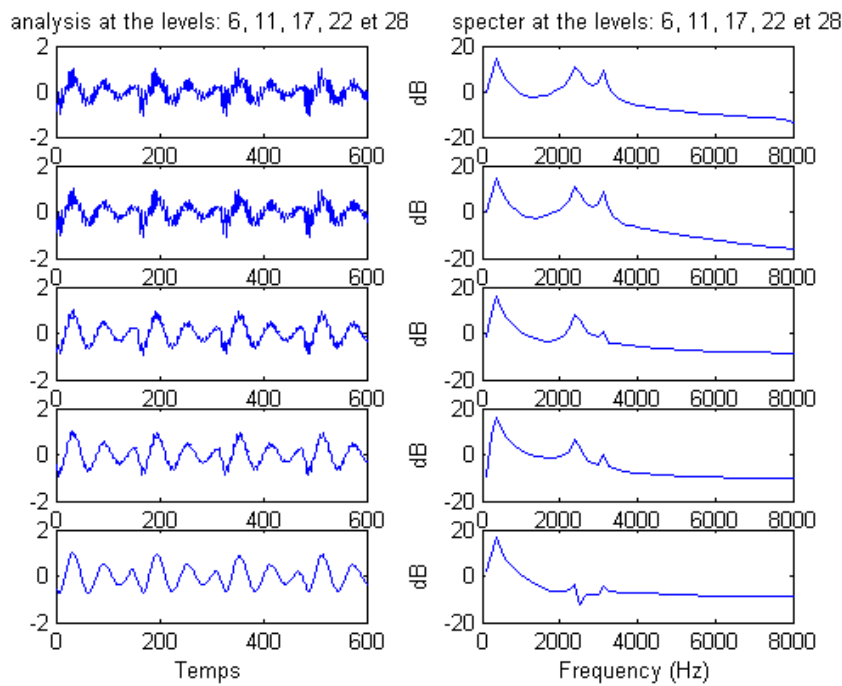

Figure 14. Application of the Gammachirp wavelet on the vowel /i/

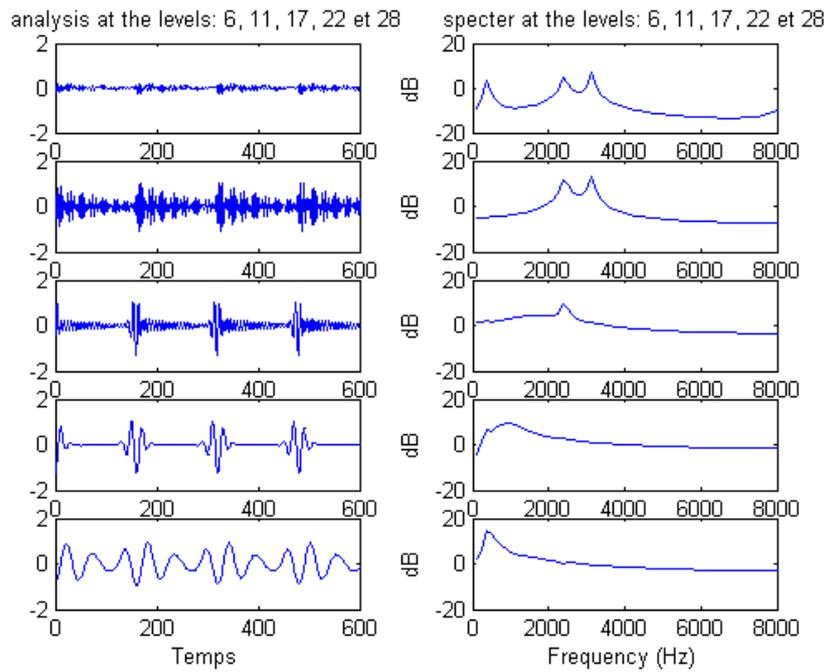

Figure 15. Application of the Morlet wavelet on the vowel /i/

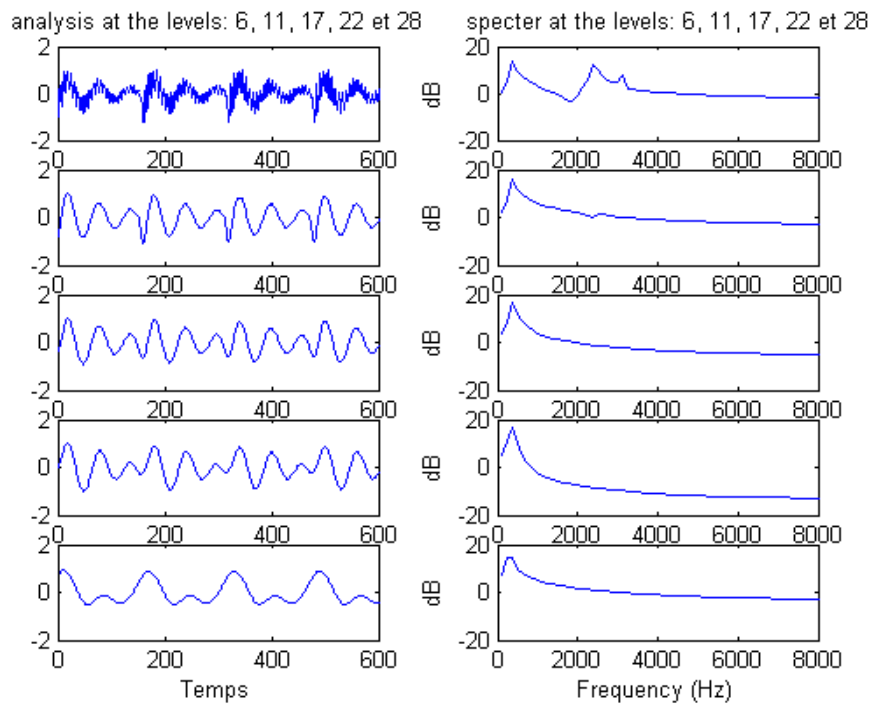

Figure 16. Application of the Mexicain Hat wavelet on the vowel /i/

Figure 14 gives a summary of the results gotten by the different wavelets:

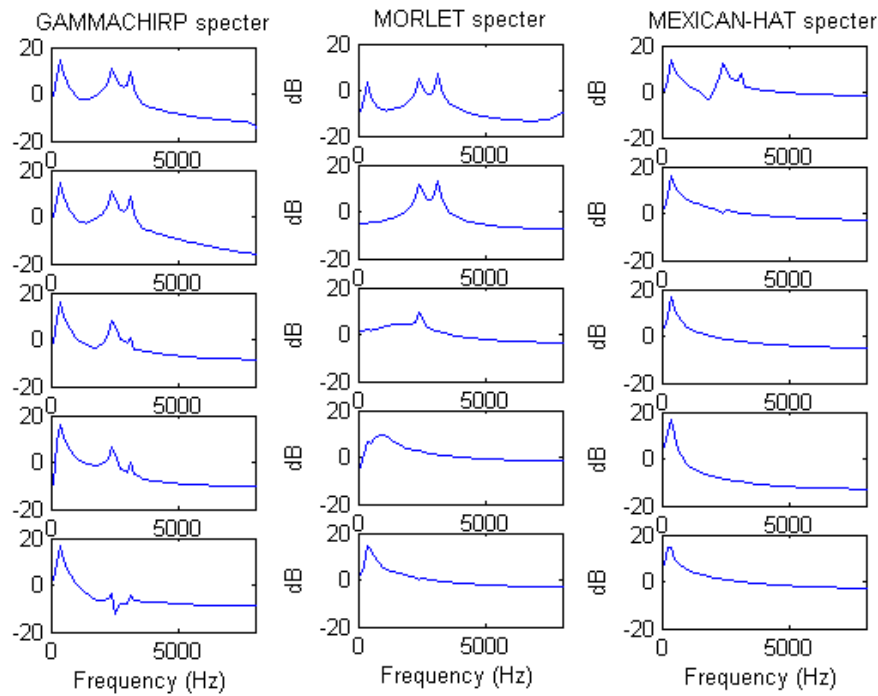

Figure 17. Comparison of the three results gotten for the vowel /i/

The first forming of the vowel /i / corresponds to the frequency 270 Hz. It is detected in the first five levels of analysis. It is waited, since the last level is limited by the frequency 522 Hz. The second forming corresponds to the frequency 2290 Hz. One notices that it is only detected in the first four levels of analysis and this while applying the Gammachirp wavelet (Figure 14). Whereas by application of Morlet (Figure 15), one notes the absence of the second forming dice the third level of analysis (Figure 16).

One recalls that for the analysis by the Gammachirp wavelet, the third level corresponds to the limit frequency 2003 Hz, whereas the fourth level corresponds to the limit frequency 1087 Hz. Although the frequency of the second forming (2290 Hz) passes these two limits, one notes its presence in these two levels. The third forming corresponds to the frequency 3010 Hz. One notices the disappearance of that forming dice the third level and this practically in the three types of wavelet.

In the case of the analysis by the gammachirp wavelet, one sometimes notices some fluctuations of the specter. These oscillations correspond to the forming of which their frequencies pass the limit frequencies of the analysis levels. It is due to the fact that the spectral slope of the wavelet is not perfectly vertical. The cut of the frequency axis in strips can have the overlaps, what explains the redundancy of the cover of the frequency axis given by this wavelet type.

Thus, with regard to the detection of the three firsts forming (Figure 17), one notices that the results found by application of the gammachirp wavelet are comparable at those given by the Mexican hatwavelet. The Morlet wavelet presents a light improvement opposite the two firsts in the distinction between the forming of the vowel /i /.

## 5. CONCLUSION

While being based on the results gotten by application of the Gammachirp wavelet on the different above-stated vowels and while comparing them at those gotten by application of other types of wavelet families, one notices that this filter gives acceptable results and that present specificities of remarkable improvement sometimes. Indeed one used the two families of Morlet wavelet and Mexican Hat wavelet for the comparison because they are in the same way standard that the Gammachirp wavelet.

Concerning this article, we presented the implementation in wavelet of the gammachirp model of the cochlear filter. We validated this implementation by its use in analysis of some vowels. The results gotten after application of this filter on the vowels /a/, /u / and /i / show that this filter gives acceptable and sometimes better results by comparison at those gotten by other types of predefined wavelet families as Morlet and Mexican hat.

**Authors**

**Lotfi Salhi** is a researcher member of the Signal Processing Laboratory in the University of Tunis - Sciences Faculty of Tunis (FST). He received his Bachelor in physics from the Sciences Faculty of Sfax (FSS). He received the diploma of Master degree in automatic and signal processing (ATS) from the National School of Engineers of Tunis (ENIT). Currently, he works a teacher of physics sciences and he prepares his doctorate thesis focused on the Analysis, Identification, and Classification of Pathological Voices using automatic speech processing.

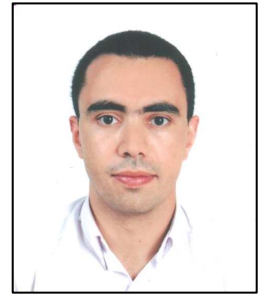

**Kais Ouni** is a Professor in the field of signal processing. He received the M.Sc. from National School of Engineers of Sfax (ENIS), the Ph.D. from National School of Engineers of Tunis (ENIT), and the HDR from the same institute. He has published more than 60 papers in Journals and Proceedings. Actually, he is a researcher member of Systems and Signal Processing Laboratory (LSTS), ENIT, Tunisia. His researches concern speech and biomedical signal processing. He is Member of the Acoustical Society of America and ISCA (International Speech Communication Association).

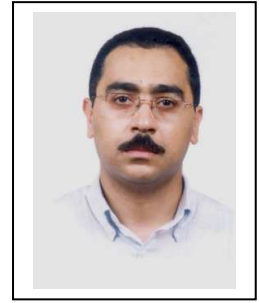